\documentclass[a4paper,fleqn]{cas-dc}

\usepackage[numbers]{natbib}

\def\tsc#1{\csdef{#1}{\textsc{\lowercase{#1}}\xspace}}
\tsc{WGM}
\tsc{QE}
\tsc{EP}
\tsc{PMS}
\tsc{BEC}
\tsc{DE}

\def\R{R}

\def\DD{{\cal D}}

\def\B{{\cal B}}

\def\P{{\cal P}}

\newcommand{\nnabla}{{\boldsymbol \nabla}}

\newcommand{\bsig}{{\boldsymbol \sigma}}
\newcommand{\bs}{{\boldsymbol \tau}}
\newcommand{\bbs}{{\boldsymbol b}}
\newcommand{\bm}{{\boldsymbol m}}
\newcommand{\bu}{{\boldsymbol v}}
\newcommand{\D}{{\boldsymbol D}}
\newcommand{\W}{{\boldsymbol W}}
\newcommand{\Rb}{{\boldsymbol R}}
\newcommand{\M}{{\boldsymbol M}}

\newcommand{\I}{{\boldsymbol I}}

\newcommand{\btt}{{\boldsymbol t}}

\newcommand{\bn}{{\boldsymbol n}}
\newcommand{\ff}{{\boldsymbol f}}

\begin{document}
\let\WriteBookmarks\relax
\def\floatpagepagefraction{1}
\def\textpagefraction{.001}
\shorttitle{Nano pillars plasticity}
\shortauthors{Salman and Ionescu}

\title [mode = title]{Tempering  the mechanical response of FCC micro-pillars: an Eulerian plasticity approach }                      

\author[1]{O\~guz Umut Salman}[orcid=0000-0003-0696-521X]

\ead{umut.salman@polytechnique.edu}

\address[1]{CNRS, LSPM UPR340, Universit\'e Sorbonne Paris Nord, Villetaneuse, France}

\address[2]{IMAR, Romanian Academy, Bucharest, Romania}

\author[1,2]{Ioan R. Ionescu}[  
orcid=0000-0003-0760-4870 ]
\cormark[1]
\ead{ioan.r.ionescu@gmail.com}


\cortext[cor1]{Corresponding author}

\begin{abstract}
The mechanical response of almost pure single-crystal micro-pillars under compression exhibits a highly localized behavior that can endanger the structural stability of a sample. Recent experiments revealed that the mechanical response of a crystal is very sensible to both  the presence of a quenched disorder in the sample, and the orientation of the crystal. In this work, we study the influence of disorder and crystal orientation on the large strain  response of a 2D FCC crystal with three  active glide planes using a very simple Eulerian plasticity model in the FE framework. Our numerical and theoretical results on clean crystal pillars suggest that a single plane or many gliding planes can be activated depending on the crystal orientation.   While in the former case, the deformation is localized, leading to ductile rupture, in the latter, a complex interplay between active planes takes place, resulting in a more uniform deformation.
The strain-localization can be avoided when inhomogeneities are engineered inside the crystal, or the crystal orientation is altered because of the activation of multiple slip systems, resulting in a "patchwork" of the distribution of the slip systems.

%
\end{abstract}

%
%

\maketitle

\section{Introduction}
When subjected to an external load, crystalline materials undergo plastic deformation beyond a material dependant elastic limit~\cite{Wilson:a28461,Salman2019-no}. At the crystal lattice level, the origin of the irreversible plastic deformation is the generation and crystal-symmetry dependent motion of linear crystal lattice defects, dislocations~\cite{hull2001introduction,Baggio2019-rs}. At large scale, the elasto-plastic mechanical response of crystalline materials manifest itself in the form of continuous strain-stress curves that rendered possible the development of continuum phenomenological theories based on some preassigned irreversible plastic flow rules when stresses exceed given thresholds \cite{Roters2010-cw,Forest2019}. Although these theories rely on the homogenisation of spatial heterogeneities such as dislocation cores, grain boundaries and mesoscale dislocations patterns etc., typically present in crystals, they are highly successful in reproducing many of aspects of plastic flow such as work-hardening, yield stress and plastic shakedown to mention some~\cite{Hughes2018-ge,Ask_undated-va,McDowell2019-jl}.

\begin{figure}
	\centering
		\includegraphics[scale=.14]{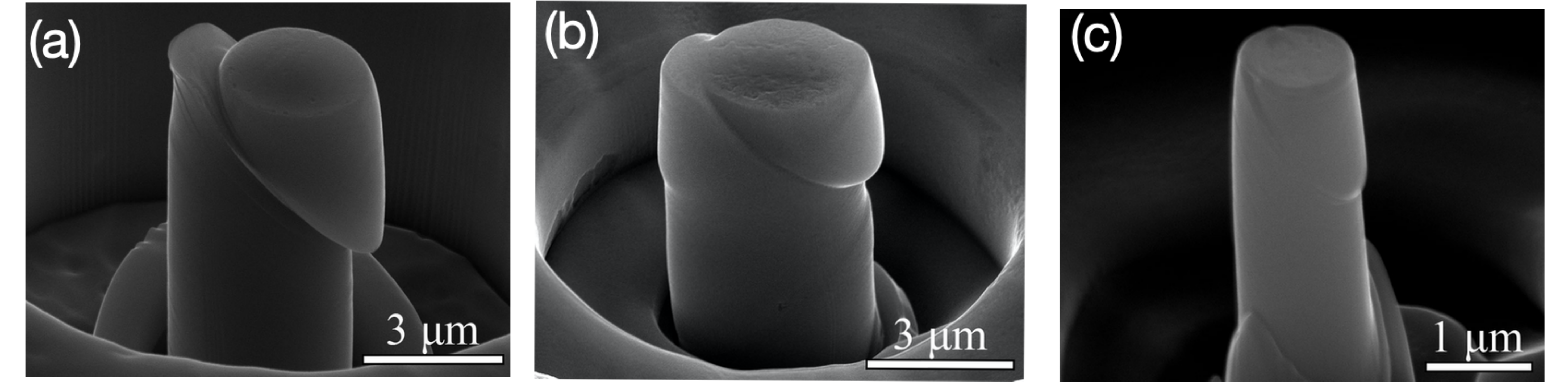}
	\caption{Scanning electron microscopy images showing the deformed states pillars: (a) the localisation of deformation  of an almost disorder-free Al crystal. Adding quenched disorder leads to a more uniform deformation as shown in (b) and (c).  Taken from \cite{Zhang2017-cl}.}
	\label{FIG:1}
\end{figure}

Our ability to control plastic flow in applications is of fundamental importance for the reliable mechanical functioning of small devices. In recent years, the single-crystal pillar compression tests became the standard tool to study mechanical response at nano and micro scales~\cite{Papanikolaou2012-up,Maas2017-yx,Sparks2018-zp,Pan2019-sl}. These experiments put in evidence unambiguously a non-smooth mechanical response with discernible stress-drops and localized deformation endangering the structural stability of the materials (see Fig. \ref{FIG:1}). This finding led to the development of new quantitative strategies in order to be able to reach a more smooth and delocalized mechanical response, highly desirable in applications. The general idea in these strategies consists in hindering dislocation motion by introducing solutes or precipitates inside the pillars leading to "dirtier is milder" effect~\cite{Zhang2017-cl,Pan2019-sl,Zhang2020-ve}. On the other hand, these strategies are not only limited to "tempering" of mechanical response but have also been used to design nanostructured crystals with ultrahigh strength and large plasticity~\cite{Wu2019-lr}. Similarly, crystal orientation has also been shown to have an important effect on the mechanical behavior in experiments such that a low-symmetry orientation results in "milder" behavior~\cite{Sparks2018-zp}. 
 
From the modeling point of view, these findings constitute a new challenge for the continuum of theories of crystal plasticity. In physical terms, the resulting overall mechanical response of pillars depends on the size, shape, and distribution of the disorder due to complex interactions of dislocations with precipitates at the lattice scale, making a continuum description difficult. However, although atomistic simulations \cite{bulatov1998connecting,baruffi2019overdamped} or mesoscopic approaches such as discrete dislocation dynamics \cite{Ispanovity2014-ra} and Landau-type theories of crystal plasticity~\cite{Salmant,Salman2011-ij,SALMAN2012219,Baggio2019-rs,zhang2020variety} can be used to describe both dislocation precipitate interactions and crystal orientation, accounting for complex mechanical loading and realistic geometries are not within the reach of even the most developed codes and can only be undertaken in the finite element framework. In this work, we study the mechanical response of a 2D FCC crystal when it is compressed in the framework of a continuum Eulerian visco-plasticity approach \cite{CI09m}. Our approach is deliberately minimalist such that the elastic properties of the lattice are neglected as they are assumed to be small in comparison with the large plastic deformations, and we used a rate-dependent (viscoplastic) model of Perzyna type without resorting to physically motivated dislocation-density based models. Since the considered rate of deformation and the viscosity are very low, the numerical results correspond to a quasi-static loading of rate-independent materials. We first consider "clean" crystals (i.e., with a homogeneous yield limit) and investigate the deformation of a crystal with respect to its orientation. Afterward, we incorporate the effects of precipitates on the deformation by assuming a higher stress flow in some pre-selected regions inside the crystal. 

The outline of the paper is as follows. We first give a brief description of the continuum  Eulerian plasticity approach developed  in  \cite{CI09m} (section 2).  In section 3  we  deduce from a  limit load analysis    some theoretical  features of slip and kink stationary shear bands.   Then we present the numerical implementation \cite{CI09n} used in the simulations (section 4). Finally, section 5  is dedicated to the presentation of the numerical results. 
%

\section{Model}

We begin by recalling from \cite{CI09m} the mechanical model used in this paper. The equations governing the motion in a domain $\DD=\DD(t)$ of an incompressible rigid-viscoplastic crystal.  Let $\bu : [0,T]\times \DD \to \R^3$, the the velocity, $\bs : [0,T] \times \DD \to \R^{3\times3}_S $,  the deviator of the Cauchy stress tensor and $p : [0,T] \times \DD \to \R$, the pressure (mean stress) ($\bsig=\bs-p\I$ is the Cauchy stress tensor) be the principal unknowns fields, while the mass density $\rho >0$
 and the body forces $\ff$ are considered known. 
The lattice orientation of the crystal is modeled through the slip direction distribution $\bbs_s: [0,T]\times \DD \to \R^3$ and the slip plane normal distribution $\bm_s: [0,T]\times \DD \to \R^3$ for each crystallographic system $s=1,2, .., N$ while the slip rate on each system $s$ will be denoted by $\dot{\gamma}_s$. 

The governing equations of the in-plane deformation of a single crystal in the domain $\DD=\Omega \times \R$ take a simpler form if we make use of only two-dimensional vector and tensor quantities. Rice \cite{rice87} showed that certain pairs of the 3-D systems that are potentially active need to combine in order to achieve plane-strain deformation. Let $r=1,..., N_{p}$ be the index of the composite slip system formed by two 3D slip systems $k$ and $l$.
%
Let $\bar{ \bbs}_k,\bar{ \bbs}_l , \bar{ \bm}_k,\bar{ \bm}_l $ denote the normalized projections of the 3D slip directions $\bbs_k, \bbs_l$ and normal directions
 $\bm_k, \bm_l$ onto the $x_1x_2$-plane, which are also orthogonal in 2D.
  We can define the in-plane slip directions for each $r=1,..N_p$ as $\bar{ \bbs}_r=\bar{ \bbs}_k$  and the in-plane plane normals as $ \bar{\bm}_r= \bar{\bm}_k$ and introduce  $\bar{\M}_r=\frac{1}{2}\left( \bar{ \bbs}_r \otimes \bar{\bm}_r+ \bar{ \bbs}_r \otimes \bar{\bm}_r\right)$ and $ \bar{\Rb}_r=\frac{1}{2}\left( \bar{ \bbs}_r \otimes \bar{\bm}_r- \bar{ \bbs}_r \otimes \bar{\bm}_r\right)$, the symmetric and antisymmetric parts of the two-dimen-\\sional tensor product of slip directions and normals.  

For plane-strain conditions let $\bar{\bs} \in \R^{2\times2}_S$ be the in-plane Cauchy stress deviator ($\bar{\tau}_{\alpha \beta}={\tau}_{\alpha \beta}$ with $\alpha,\beta=1,2$) and  let $\bar{\bu}\in \R^2$ be the in-plane velocity ($\bu=(\bar{\bu},0), \; \bar{\bu}=\bar{\bu}(t,x_1,x_2)$). Let also denote by $\displaystyle \D(\bar{\bu}):= \frac{1}{2}(\nnabla \bar{\bu}+\nnabla^t \bar{\bu})$ and by $\displaystyle \W(\bar{\bu}):= \frac{1}{2}(\nnabla \bar{\bu}-\nnabla^t \bar{\bu})$, the symmetric and antisymmetric parts, respectively, of the two-dimensional velocity gradient $\nnabla \bar{\bu}$.


In applications involving large deformations and high strain rates, the elastic component of the deformation is small with respect to the inelastic one. That is why it can be  neglected it here and a rigid-viscoplastic approach adopted (e.g. \cite{hut76, lebtome93,kok02}). 
As it is proven in \cite{CI09m},  the two-dimensional rate of deformation tensor $\D( \bar{\bu})$ can be decomposed into $N_p$ composite plane-strain systems specified by $(\bar{\bbs}_r, \bar{\bm}_r)$, the normalized projections of the three-dimensional slip directions and normal directions onto the plane. Thus, 
\begin{equation}\label{Dg} \displaystyle 
 \D(\bar{\bu})= \sum_{r=1}^{N_p}\dot{\bar{\gamma}}_{r}\bar{\M}_r, 
 \end{equation}
where $\dot{\bar{\gamma}}_r=2q_r\dot{\gamma}_r$, and $q_r$ are the in plane factors specific to each crystal type, we get the in-plane form of the flow rule 
\begin{equation}\label{flow_rule2D}
 \dot{\bar{\gamma}}_r = \dfrac{1}{\bar{\eta}_r}\left[1-\dfrac{\bar{\tau}_{r}^c}{\vert \bar{\bs} : \bar{\M}_r \vert}\right]_+ \bar{\bs} : \bar{\M}_r,
\end{equation}
where $\bar{\eta}_r=\eta_r/(2q_r^2)$ and $\bar{\tau}_{r}^c=\tau_{r}^c/q_r$ denote the "in-plane" viscosity and yield limit.

If we specify the two-dimensional vectors $ \bar{\bbs}_r$ and $ \bar{\bm}_r$ by their polar coordinates
$ \bar{\bbs}_r=(\cos\theta_r, \sin\theta_r), \quad \bar{\bm}_r = (-\sin\theta_r, \cos\theta_r)$,
then  the  angles between two systems, say $\theta_r - \theta_q$, do not change in time, and it is sufficient to compute the change in orientation of only one of the composite in-plane slip systems, for instance $\theta=\theta_1$,  by using the equation
\begin{equation}\label{theta}
 \displaystyle \frac{\partial \theta}{\partial t} + \bar{\bu}\cdot \nabla \theta= \frac{1}{2}\left( \sum_{r=1}^{N_p}\dot{\bar{\gamma}}_{r} -(\frac{ \partial v_1}{ \partial x_2}-
 \frac{ \partial v_2}{ \partial x_1})\right).
\end{equation}
Then the orientation of f all other $N_p-1$ composite plane strain systems can be obtained from the relation
 $\theta_r(t)= \theta(t)+\theta_r(0)-\theta_1(0)$.
 
 The yield limits $\bar{\tau}_{r}^c$ of each slipping system $r$ can be considered as constants, but they can vary in time if hardening effects are taken into consideration. Since the slip on each system produces hardening on all slip systems, the slip resistances $\bar{\tau}_{r}^c$ have an Eulerian evolution law of the form:
\begin{equation}\label{hardening-law}
\partial_t \bar{\tau}_{r}^c + \bar{\bu}\cdot \nabla\bar{\tau}_{r}^c= \sum_{s=1}^{N} h_{sr}\vert \dot{\bar{\gamma}}_{s}\vert. 
  \end{equation}
 The matrix $h_{sr}$, called hardening matrix, describes the slip resistance on system $s$ which is caused by slip on system $r$. As shown by Franciosi \cite{fra85}, the matrix $h_{sr}$ is not constant. Expressions for the hardening matrix, $h_{sr}=h_{sr}(\gamma)$, on the cumulated shears on all systems 
 are widely used in FE simulations of polycrystals (see \cite{peir82,anakoth96}). Hardening laws that use dislocation densities on all slip planes as internal variables have also been proposed (e.g. \cite{TeodSid76,TeodRafTab}).

Let us see now how the in-plane model is applied at an FCC crystal, which corresponds to the case used in the next sections. For a FCC crystal, which has  12 potentially active 3-D slip systems,  let $Ox_3$ axis be parallel to $ [1 1 0] $ in the crystal basis, which means that  the plane-strain plane $Ox_1x_2$ is  the plane $[\bar{1}10]-[001]$.   
To describe the orientation of the crystal we denote by $\theta$ the angle counterclockwise between the $Ox_1$ axis and $[\bar{1}10]$ direction. The three active composite in-plane slip systems $\bar{ \bbs}_1, \bar{ \bbs}_2, \bar{ \bbs}_3$ will be specified by the angles  $
\theta_1 =\theta, $ $\theta_2 = \theta+\phi,$ and $\theta_3 = \theta-\phi$, with $ \phi=\arctan(\sqrt{2})$,  while the corresponding in-plane factors are $ q_1=1/\sqrt{3}, q_2=q_3= \sqrt{3}/2$.  
 
 \section{Stationary shear bands in a FCC crystal}

 We shall try to understand how the in-plane model, presented in the previous section, can describe the mechanism of stationary shear bands in "clean" FCC crystals. For simplicity, we will suppose that the slip resistance for all systems is equal to $\tau^c$, and no hardening is taken into consideration. This means that the in-plane yield limit for each systems, $\bar{\tau}^c_1=\sqrt{3}\tau^c$ and $\bar{\tau}^c_2=\bar{\tau}^c_3=2\tau^c/\sqrt{3}$ are constant. Moreover we consider a vanishing viscosity ($\bar{\eta}_r=0$), i.e., we deal with a rigid-perfectly plastic model. 
  
\begin{figure}
	\centering
		\includegraphics[scale=.21]{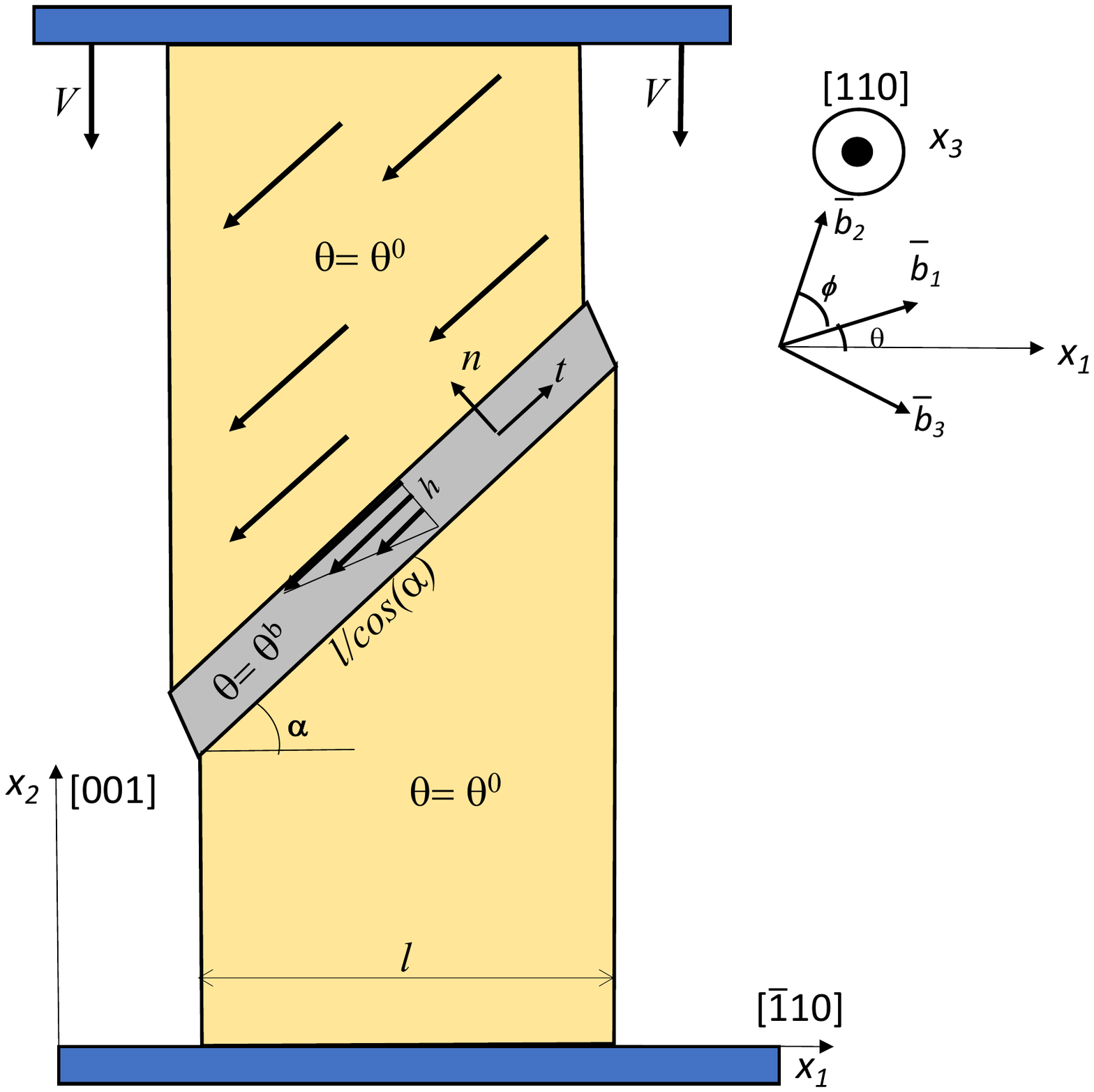}
	\caption{A schematic representation of a shear band  in a rigid-perfectly plastic  model.}
	\label{FigBand}
\end{figure}

The approach used here to describe a shear band is not following Assaro and Rice (see \cite{Asaro1977-nn} but also \cite{Asaro2006-uk,Marano2019-he,Borja2001-hr}) for elastic-plastic models with hardening. For our simple rigid-plastic model, another method, used in modeling ductile rupture, called limit load analysis, seems to be more appropriate. To model shear bands, where strains are localized on some surfaces, the associate (collapse) velocity has to exhibit discontinuities. The block decomposition method, which was intensively used in the analytical developments (see  for instance 
\cite{Hill1998-ae,Kachanov1971-he}), consists in considering only velocity fields generated by a decomposition of the structure in rigid blocks separated by discontinuity surfaces. The plastic dissipation power is then minimized on this particular class of functions (see also \cite{Ionescu2010-zc}). 
The main difference between the present model and a classical plasticity approach of the block decomposition method is the role played by the evolution of the crystal orientation $\theta^b$ in the shear band through (\ref{theta}).  

The "clean" crystal, occupying the domain $(0,l)\times (0,H)$ and having the homogeneous orientation $\theta=\theta^0$, is splinted into two rigid regions (see Fig. \ref{FigBand}) separated by a shear band in the direction $\btt=(\cos(\alpha),\sin(\alpha))$. The upper region is animated by a velocity $-V/sin(\alpha)\btt$, where $V$ is the vertical velocity of the upper plateau, while the bottom region is at rest. In the shear band region, denoted by $\B$ and having the width $h$, the crystal has the orientation $\theta=\theta^b$. The associated velocity field $\bar{\bu}$ has the strain rate $ \D(\bar{\bu})=-V/(h\sin(\alpha))\btt\otimes\bn I_\B$, where $I_\B$ is the indicator function of the shear band $\B$ ($I_\B(x)=1$ if $x\in\B$ and $I_\B(x)=0$ if not).  

The slip rates on each system in the shear band region $\dot{\bar{\gamma}}_{r}^b$, $r=1,2,3$ can  be determined by minimizing the internal plastic dissipation power $ J(\dot{\bar{\gamma}}_{1}, \dot{\bar{\gamma}}_{2},\dot{\bar{\gamma}}_{3})= \sum_{r=1}^{3}  \bar{\tau}_r^c |\dot{\bar{\gamma}}_{r}|,$ under the kinematic
constraints (\ref{Dg}). Moreover, we can use the closed form from  \cite{CI09m} to find the  analytic expressions of  $\dot{\bar{\gamma}}_{r}$ for any crystal orientation $\theta^b$. The evolution of the crystal orientation in the shear band can be obtained from the differential equation (\ref{theta}), which reads $\dot{\theta}^b=[\dot{\bar{\gamma}}_{1}^b+ \dot{\bar{\gamma}}_{2}^b+\dot{\bar{\gamma}}_{3}^b+V/(h\sin(\alpha))]/2$. For stationary (permanent) shear bands the  crystal orientation in the shear band is constant in time, hence we have a supplementary equation for the slip rates $$\dot{\bar{\gamma}}_{1}^b+ \dot{\bar{\gamma}}_{2}^b+\dot{\bar{\gamma}}_{3}^b=-\frac{V}{h\sin(\alpha)}.$$ 
  Using this last equation we can prove that {\em there exists only three possible orientations of the crystal in a stationary shear band $\theta^b=\alpha, \alpha-\phi,  \alpha+\phi$ and for each orientation there exists only one active slipping system $\dot{\bar{\gamma}}_r^{b}\neq 0$}.    
  
 Let us analyse now the link between the crystal orientation in the shear band $\theta^b$ and on the clean crystal $\theta^0$.  We have to check that the requirement of continuing equilibrium $\bar{\bs}^b\bn\cdot\btt= \bar{\bs}^0\bn\cdot\btt$ is compatible with the rigid-perfect plastic law  $|\bar{\bs}^0\bar{ \bbs}_r\cdot\bar{ \bm}_r |\leq \bar{\tau}_c^r$ for all $r$. We found that if $\theta^b=\alpha$ then there exists a restriction on $\theta^0$ but there are no restrictions for $\theta^b=\alpha \pm \phi$. The case $\theta^0=\theta^b$ is always possible and following \cite{Marano2019-he}, we called it {\em slip band} while for $\theta^0\neq\theta^b$ we called it {\em kink band}.  Having in mind the link between the shear band angle and the orientation of the crystal in the shear band we obtain that the {\em slip bands are possible only for a special orientation of the cystal} $\theta^0=\alpha, \alpha-\phi,  \alpha+\phi$.  

The total plastic dissipation power associated to the velocity field $\bar{\bu}$, $\P(\bar{\bu})= \sum_{r=1}^{3}  \int_\B \bar{\tau}_r^c |\dot{\bar{\gamma}}_{r}^b |$, is then given by $ \P=2V\bar{\tau}^c_1l/ |\sin(2\alpha) | $ if $ \theta^b=\alpha$ and $\P=2V\bar{\tau}^c_2 l/|\sin(2\alpha) |$ if $\theta^b=\alpha \pm \phi$. Moreover the plastic dissipation power is independent of the shear band thickness $h$.  That is why the shear band can be considered with a vanishing width, which corresponds to a discontinuous velocity field. We expect that a stationary band is stable if the total plastic dissipation power is minimum. Since $\bar{\tau}^c_1>\bar{\tau}^c_2=\bar{\tau}^c_3$ we deduce that the orientation in the shear band is $\theta^b=\alpha \pm \phi$. The angle $\alpha$ of the shear band which corresponds to the minimum of the plastic dissipation power $ \P$, is $\alpha=\pm \pi/4$ and $\P_{min}=2Vl\bar{\tau}^c_2$. In conclusion we expect that {\em a stable stationary shear band has an angle of $\alpha=\pm \pi/4$ while the crystal orientation in the shear band is $\theta^b= \pm \pi/4+ \phi$ or $\theta^b= \pm \pi/4-\phi$}. 

Finally, let us mention that the above analysis tells us when the shear bands can be expected, but it does not guarantee the existence of them. Moreover, the Eulerian configuration considered here has an idealized geometry corresponding to the beginning of the shear band formation. As we can see from the numerical simulations after large strains, these configurations are no more realistic, and the analysis became much more complicated.  
  
\section{Numerical approach}
We recall from \cite{CI09n} the principal features of the numerical scheme used in this paper. The time implicit (backward) Euler scheme for time discretization gives a set of nonlinear equations for the velocities $\bu$ and lattice orientation $(\bbs_s, \bm_s)$. At each time iteration,  an iterative algorithm is developed to solve these nonlinear equations. Specifically, a mixed finite-element and Galerkin discontinuous strategy are proposed. The variational formulation for the velocity field is discretized using the finite element method, while a Galerkin discontinuous method with an upwind choice of the flux is adopted for solving the hyperbolic equations that describe the evolution of the lattice orientation. It is to be noted that in the case of the rigid-viscoplastic model, additional difficulties arise from the non-differentiability of the plastic terms.  To overcome these difficulties, we have used a modified version of the iterative decomposition-coordination formulation coupled with the augmented Lagrangian method (introduced in 
\cite{GlLT}). 
The adopted visco-plastic model contains as a limit case the inviscid Schmid law. For low viscosities and moderate strain rates, the iterative decomposition coordination formulation coupled with the augmented Lagrangian method works very well, and no instabilities are present. 
 If the computational Eulerian evolves in time (see the last section), an ALE (Arbitrary Eulerian-Lagrangian) description of the crystal evolution has to be implemented.  
\section{Numerical simulations}
\begin{figure}
	\centering
		\includegraphics[scale=.29]{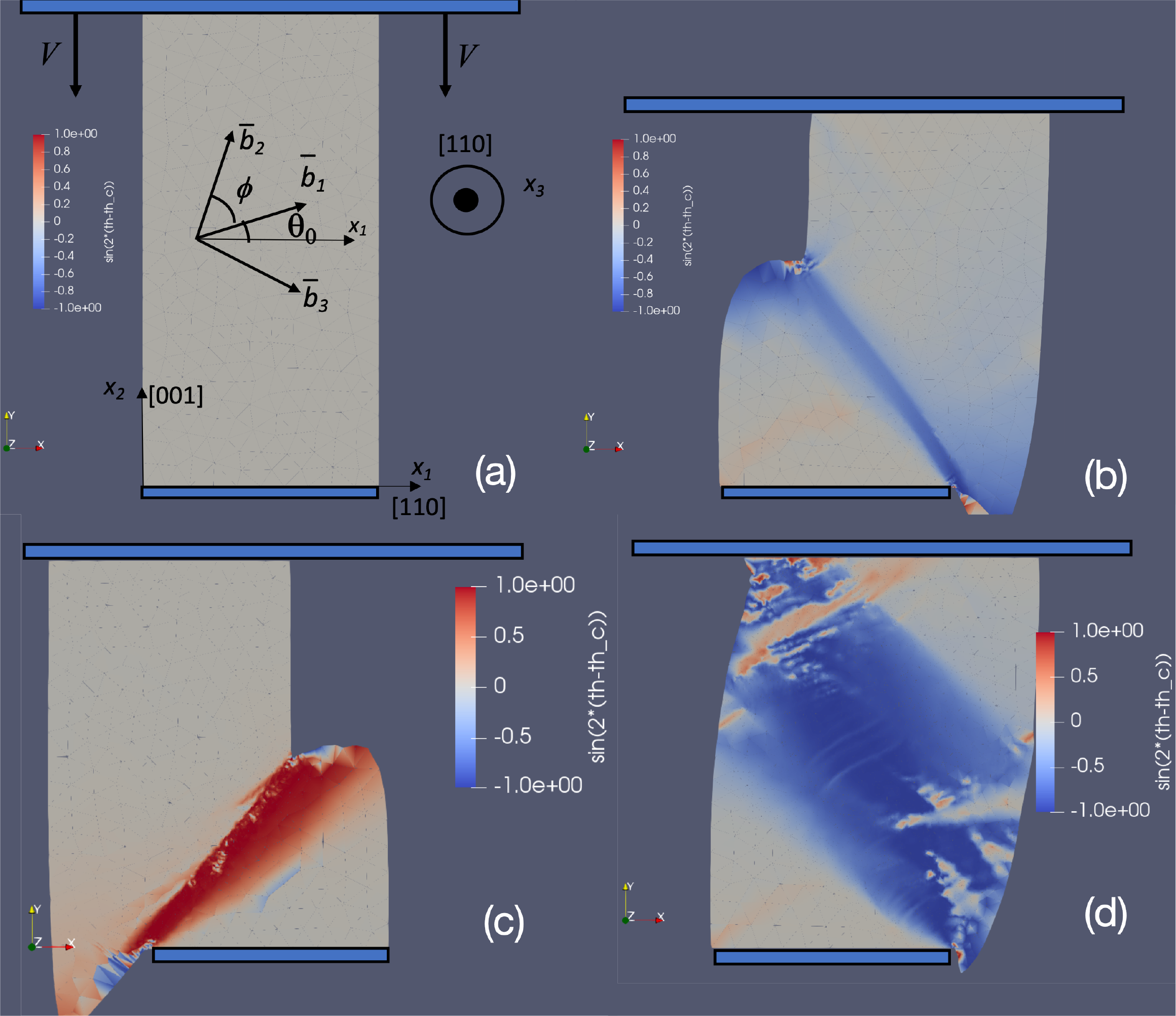}
	\caption{(a): The initial geometry of a "clean" single crystal with a homogeneous initial crystal orientation $\theta^0$ in the plane strain compression configuration. The final states of "clean" crystals with different initial orientation $\theta^0$:  (b) $\theta^0=\pi/2$, (c) $\theta^0=20^o$, (d) $\theta^0=0$. The distribution of the orientation misfit $\sin(2(\theta-\theta^0))$ is plotted in a color scale.}
	\label{FIG:2}
\end{figure}

In this section, we analyze the in-plane compression of a micro-pillar. The micro-pillar is modeled as an FCC crystal with $ [1 1 0] $ axis of the crystal is along $Ox_3$ (see Fig. \ref{FIG:2}(a) for a schematic representation and the end of section 2 for more details). 
 The material coefficients considered correspond to tantalum (Ta) of density $\rho=16,650$Kg/m$^3$ and of slip resistance for all systems equal to $\tau_0^s=66$MPa (i.e. $\bar{\tau}_0^1=114.31$MPa, $\bar{\tau}_0^2=\bar{\tau}_0^3=76.23$MPa) and  no hardening is considered.  
 In all the computations, the initial configuration of the micro-pillar has a rectangular section of $l\times H$, with $l=$1$\mu$m, $H=$2$\mu$m.  The rate of deformation is very low (10$^{-4}$s$^{-1}$) such that the loading could be considered as quasi-static and the time interval $[0,T]$ was chosen such that the final height corresponds to a quarter of the initial height of the sample, i.e., a $25\%$ reduction.  
 
 To capture the shear bands, we have used an adaptive mesh technique with respect to the strain rate norm $\vert  \D(\bar{\bu}) \vert$. That means that the regions where the slip rate is larger will have a fine mesh while outside the mesh is coarse. The ratio between the sizes of the fine and coarse mesh was $1/8$. Since we deal with an implicit numerical scheme, the chosen time step is large and corresponds to a deformation of $0.1\%$  between two steps in time. That is why the computational cost is low, and simulations can be performed on a personal computer. 

Firstly we investigate the compression of a "clean" homogeneous crystal, i.e., with a uniform yield limit and initial orientation $\theta=\theta^0$. In Fig. \ref{FIG:2}, we show the final configurations of the pillars for three different initial orientations. We remark that the numerical scheme was able to reproduce the shear bands and the ductile rupture associated with it~\cite{Zhang2020-ve}.  Note that capturing a discontinuous phenomenon, such as rupture,  is a  challenging  task for  a continuous FE technique.   We distinguish two types of deformation: (i) a thin shear band separating two rigid regions (as in (b) and (c)) and (ii) a more diffuse deformation configuration as in (d).

\begin{figure}
	\centering
		\includegraphics[scale=.12]{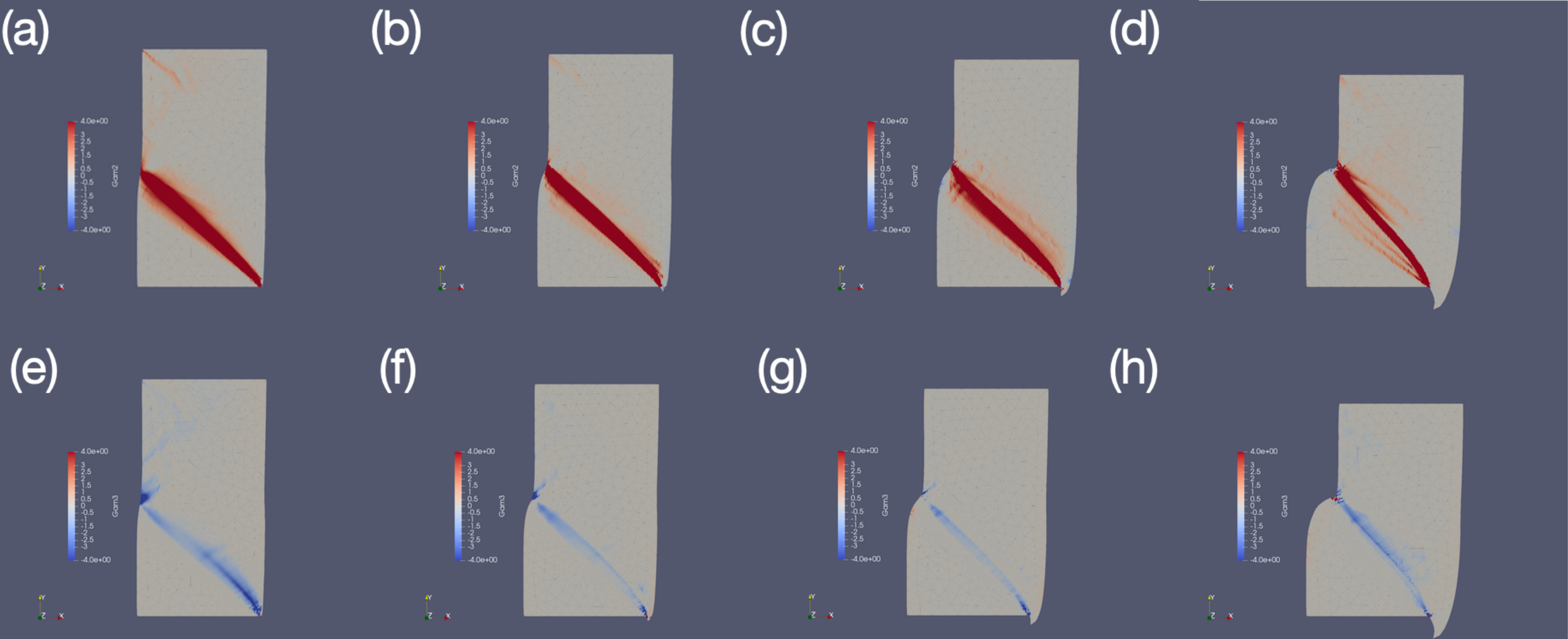}
	\caption{The slip rates $\dot{\bar{\gamma}}_2$ (up) and $\dot{\bar{\gamma}}_3$ (bottom)  distribution for a "clean" crystal with initial orientation $\theta^0=\pi/2$ at four different levels of deformation.  }
	\label{FIG:3}
\end{figure}

 (i) As it is predicted by the simple model given in section 3, the shear band deformation mechanism is related to a single active slipping system while the other two are inactive. To see that we have plotted in Fig. \ref{FIG:3} the slip rate $\dot{\bar{\gamma}}_2$ and $\dot{\bar{\gamma}}_3$ for the pillar with the initial orientation $\theta_0=\pi/2$ (corresponding to Fig. \ref{FIG:2}(b)) at four different levels of deformation. The slip rate $\dot{\bar{\gamma}}_1$ was not plotted because this system is completely inactive. We remark that during all the compression process, the slipping system $r=2$ is very active in the shear band, while the slipping system $r=3$, which was active at the beginning, is almost inactive at the end. From the orientation distribution, plotted in a color scale in Fig. \ref{FIG:2}, we see that at the end of the deformation process the crystal has two "rigid regions" where the pillar has the same orientation as before the compression, separated by a "shear band region" where the orientation is different. In both cases, we deal with a kink band. Let analyze now the orientation, $\alpha$ of the shear band. In the case (b), the band is not exactly a straight line, and its overall angle varies in time between $45^o$ and $55^o$, not far from $\alpha= \pi/4$, predicted by the theoretical considerations of section 3.  In the case (c), the shear band is a straight line perfectly orientated at  $\pi/4$.  
 
\begin{figure}
	\centering
		\includegraphics[scale=.12]{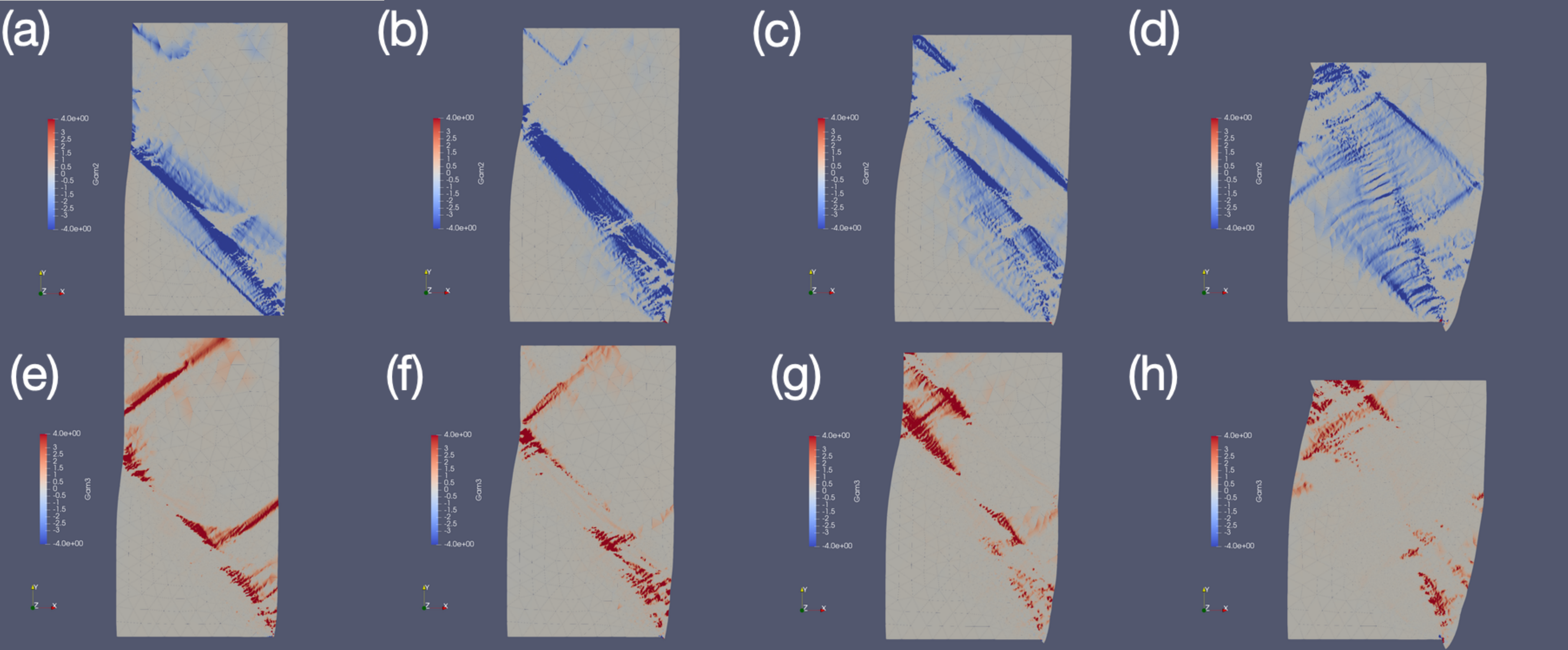}
	\caption{The slip rates $\dot{\bar{\gamma}}_2$ (up) and $\dot{\bar{\gamma}}_3$ (bottom) distribution for a "clean" crystal with  initial  orientation $\theta^0=0$ at four different levels of deformation. }
	\label{FIG:4}
\end{figure}

(ii) In the third case, corresponding to an initial orientation of $\theta^0=0^o$, the deformation mechanism is quite different. In this case two slipping systems $r=2$ and $r=3$ are active while the system $r=1$ is inactive (i.e., $\dot{\bar{\gamma}}_1=0$). We have plotted in Fig. \ref{FIG:4} two snapshots of the slip rate $\dot{\gamma}_r$ distribution, corresponding to directions $r=2$ and $r=3$  at four levels of compression. We remark that there exists a large (band) region where the deformation is not vanishing, between two small rigid regions. This deformation region is split into several regions where one of the slipping system $r=2$ or $r=3$ is active, but the two systems are not active in the same place, forming a "patchwork" of the distribution of the slipping systems. Each active slipping system is related to some specific shear bands, oriented at an angle around $\pi/4$, but we do not deal with one single shear band, which accumulates all the deformation. These multiple shear bands give a diffuse overall deformation. Concerning the crystal orientation, as we can see in \ref{FIG:2} (d) that the pillar seems to have three regions of orientations: two with the initial orientation $\theta^0$ in the rigid regions and another one in the deformation region, which could be assimilated to a kink band. However, the orientation in this last region is not uniform, with multiple bands of initial orientation.  

\begin{figure}
	\centering
		\includegraphics[scale=.15]{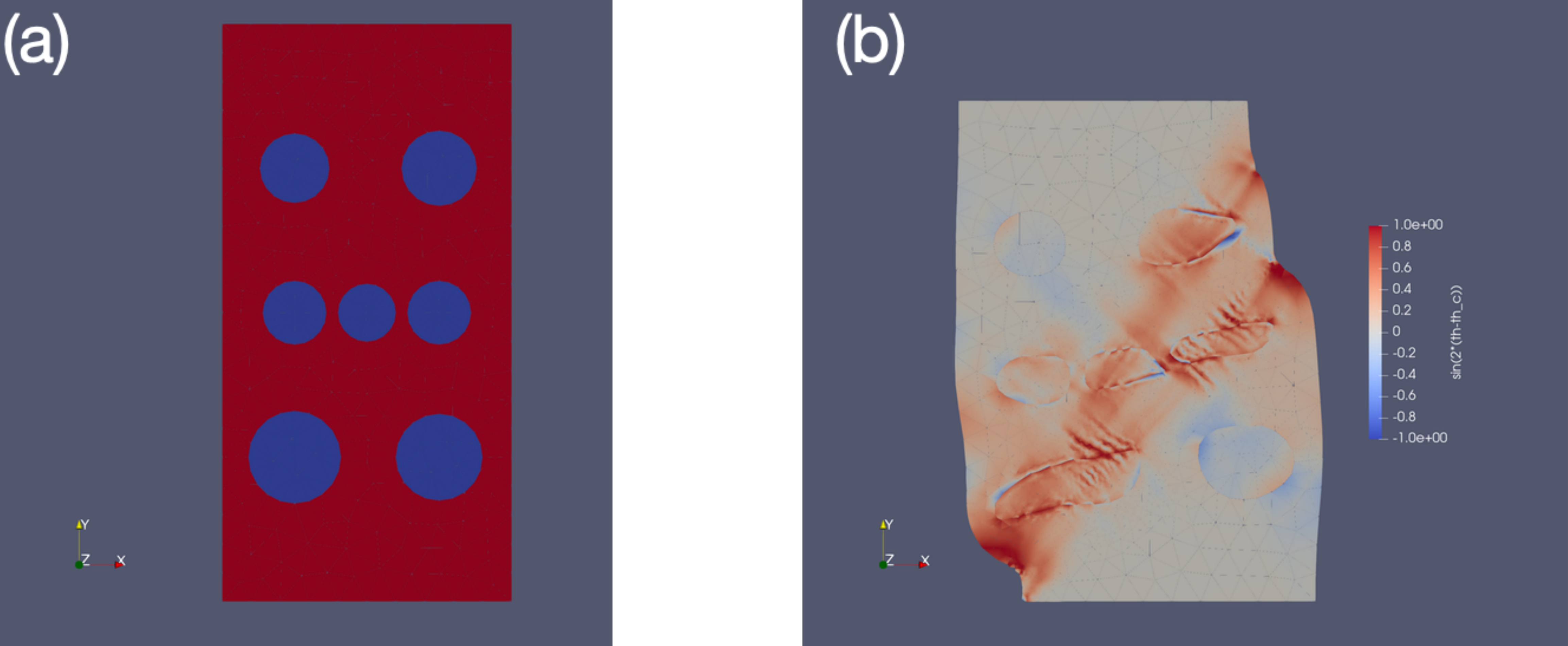}
	\caption{Initial (a) and final (b) states of  the crystal with inhomogeneities (shown in blue in (a)) in which  plastic yield threshold  is smaller. The distribution of the orientation misfit $\sin(2(\theta-\theta^0))$ is plotted  in  a color scale.}
	\label{FIG:6}
\end{figure}

We now turn our focus into "non-clean" crystals with some pre-existent inhomogeneities. This case can be considered as a size-related effect since, generally speaking, with increasing size, it becomes harder to manufacture "clean" crystals~\cite{Weiss2019-sg}. However, it is also possible to engineer the disorder into the small crystals~\cite{Zhang2017-cl}. In this work, we use some circular regions with a lower yield limit into the computational domain as a proxy for "non-clean" crystals, see Fig 6(a). For simplicity,  the orientation of these seven defects was taken to be the same as in the crystal $\theta^0$.  A more realistic approach would be, of course, to use dislocation-density based models that we will consider in future work.

For the study of the effect of inhomogeneities on macroscopic deformation, we choose the crystal orientation $\theta^0=\pi/2$ as in this case, we observed a strong localization of deformation, see Fig. 3(b). The final deformation state for the inhomogeneous case is shown in Fig. 6(b), where we observe that the kink band formation is suppressed in favor of a more uniform deformation state. To elucidate this behavior, we show in Fig. 7, the evolution of the spatial distribution of deformation rates $\dot\gamma_2$ (upper row) and $\dot\gamma_3$ (lower row), where we observe that both are steadily active during all the deformation history. Note that we did not include $\dot\gamma_3$ since it vanished everywhere almost all the time.

 \begin{figure}
	\centering
		\includegraphics[scale=.12]{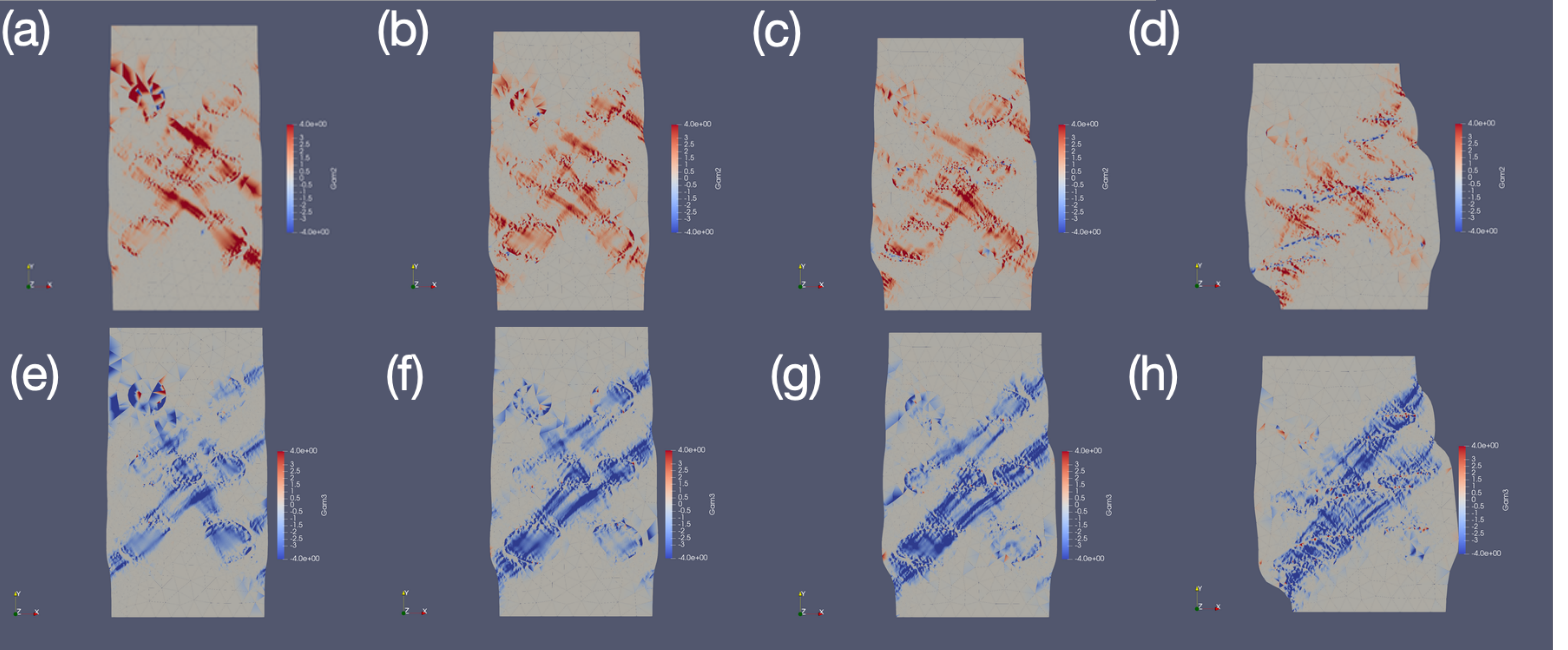}
	\caption{The slip rates  $\dot\gamma_2$ (up) and $\dot\gamma_3$ (bottom)  spatial distribution in a "non-clean" crystal at four different levels of deformation.}
	\label{FIG:5}
\end{figure}
\section{Conclusions}
 We studied the in-plane compression of micro-pillars using a "minimal" two-dimensional Eulerian plasticity approach.  The model shows a very strong localization of deformation, and we found a qualitative agreement with experiments.  From theoretical considerations, based on the rigid block decomposition method of the associated Eulerian limit load problem, we expect that the stationary kink bands are oriented at $\pi/4$ form the compression axis. Numerical simulations on clean single crystal pillar, which partially confirm the theory, show that the {compression process is not stable with respect to the initial orientation}, and it is very difficult to predict the final shape of the pillar. However, the {principal mechanism of deformation seems to be the kink shear band which separate two rigid blocks}, associated with one single active slip system. If two possible slip systems are active, then the deformation is diffuse but localized in a large width shear band.  Besides, our results show that strain-localization can be avoided when inhomogeneities are engineered inside the crystal, or the crystal orientation is altered. This is because of the activation of multiple slip systems, resulting in a "patchwork" of the distribution of the slip systems. The natural extension of our work will be to consider three-dimensional systems and incorporate dislocation-density based constitutive rules.
 
 {\bf Acknowledgements. }  This work was partially supported by the grants of the French Research Agency ANR-17-CE24-0027 and  ANR-19-CE08-0010-01.
\bibliographystyle{cas-model2-names}

\bibliography{cas-refs}

\begin{thebibliography}{43}
\expandafter\ifx\csname natexlab\endcsname\relax\def\natexlab#1{#1}\fi
\providecommand{\url}[1]{\texttt{#1}}
\providecommand{\href}[2]{#2}
\providecommand{\path}[1]{#1}
\providecommand{\DOIprefix}{doi:}
\providecommand{\ArXivprefix}{arXiv:}
\providecommand{\URLprefix}{URL: }
\providecommand{\Pubmedprefix}{pmid:}
\providecommand{\doi}[1]{\href{http://dx.doi.org/#1}{\path{#1}}}
\providecommand{\Pubmed}[1]{\href{pmid:#1}{\path{#1}}}
\providecommand{\bibinfo}[2]{#2}
\ifx\xfnm\relax \def\xfnm[#1]{\unskip,\space#1}\fi
\bibitem[{Anand and Kothari(1996)}]{anakoth96}
\bibinfo{author}{Anand, L.}, \bibinfo{author}{Kothari, M.},
  \bibinfo{year}{1996}.
\newblock \bibinfo{title}{A computational procedure for rate-independent
  crystal plasticity}.
\newblock \bibinfo{journal}{J. Mech. Phys. Solids} \bibinfo{volume}{44},
  \bibinfo{pages}{525--558}.
\bibitem[{Asaro and Lubarda(2006)}]{Asaro2006-uk}
\bibinfo{author}{Asaro, R.}, \bibinfo{author}{Lubarda, V.},
  \bibinfo{year}{2006}.
\newblock \bibinfo{title}{Mechanics of Solids and Materials}.
\newblock \bibinfo{publisher}{Cambridge University Press}.
\bibitem[{Asaro and Rice(1977)}]{Asaro1977-nn}
\bibinfo{author}{Asaro, R.J.}, \bibinfo{author}{Rice, J.R.},
  \bibinfo{year}{1977}.
\newblock \bibinfo{title}{Strain localization in ductile single crystals}.
\newblock \bibinfo{journal}{J. Mech. Phys. Solids} \bibinfo{volume}{25},
  \bibinfo{pages}{309--338}.
\bibitem[{Ask et~al.(2018)Ask, Forest, Appolaire, Ammar and
  Salman}]{Ask_undated-va}
\bibinfo{author}{Ask, A.}, \bibinfo{author}{Forest, S.},
  \bibinfo{author}{Appolaire, B.}, \bibinfo{author}{Ammar, K.},
  \bibinfo{author}{Salman, O.U.}, \bibinfo{year}{2018}.
\newblock \bibinfo{title}{A cosserat crystal plasticity and phase field theory
  for grain boundary migration}.
\newblock \bibinfo{journal}{Journal of the Mechanics and Physics of Solids}
  \bibinfo{volume}{115}, \bibinfo{pages}{167--194}.
\bibitem[{Baggio et~al.(2019)Baggio, Arbib, Biscari, Conti, Truskinovsky,
  Zanzotto and Salman}]{Baggio2019-rs}
\bibinfo{author}{Baggio, R.}, \bibinfo{author}{Arbib, E.},
  \bibinfo{author}{Biscari, P.}, \bibinfo{author}{Conti, S.},
  \bibinfo{author}{Truskinovsky, L.}, \bibinfo{author}{Zanzotto, G.},
  \bibinfo{author}{Salman, O.U.}, \bibinfo{year}{2019}.
\newblock \bibinfo{title}{{Landau-Type} theory of planar crystal plasticity}.
\newblock \bibinfo{journal}{Phys. Rev. Lett.} \bibinfo{volume}{123},
  \bibinfo{pages}{205501}.
\bibitem[{Baruffi et~al.(2019)Baruffi, Finel, Le~Bouar, Bacroix and
  Salman}]{baruffi2019overdamped}
\bibinfo{author}{Baruffi, C.}, \bibinfo{author}{Finel, A.},
  \bibinfo{author}{Le~Bouar, Y.}, \bibinfo{author}{Bacroix, B.},
  \bibinfo{author}{Salman, O.U.}, \bibinfo{year}{2019}.
\newblock \bibinfo{title}{Overdamped langevin dynamics simulations of grain
  boundary motion}.
\newblock \bibinfo{journal}{Materials Theory} \bibinfo{volume}{3},
  \bibinfo{pages}{4}.
\bibitem[{Borja(2001)}]{Borja2001-hr}
\bibinfo{author}{Borja, R.}, \bibinfo{year}{2001}.
\newblock \bibinfo{title}{Bifurcation of elastoplastic solids to shear band
  mode at finite strain}.
\newblock \bibinfo{journal}{Comput. Methods Appl. Mech. Eng.}
  \bibinfo{volume}{191}, \bibinfo{pages}{5287--5314}.
\bibitem[{Bulatov et~al.(1998)Bulatov, Abraham, Kubin, Devincre and
  Yip}]{bulatov1998connecting}
\bibinfo{author}{Bulatov, V.}, \bibinfo{author}{Abraham, F.F.},
  \bibinfo{author}{Kubin, L.}, \bibinfo{author}{Devincre, B.},
  \bibinfo{author}{Yip, S.}, \bibinfo{year}{1998}.
\newblock \bibinfo{title}{Connecting atomistic and mesoscale simulations of
  crystal plasticity}.
\newblock \bibinfo{journal}{Nature} \bibinfo{volume}{391},
  \bibinfo{pages}{669--672}.
\bibitem[{Cazacu and Ionescu.(2010a)}]{CI09n}
\bibinfo{author}{Cazacu, O.}, \bibinfo{author}{Ionescu., I.R.},
  \bibinfo{year}{2010}a.
\newblock \bibinfo{title}{Augmented lagrangian method for eulerian modeling of
  viscoplastic crystals}.
\newblock \bibinfo{journal}{Computer Methods in Appl. Mech. and Engng.}
  \bibinfo{volume}{199}, \bibinfo{pages}{68--699}.
\bibitem[{Cazacu and Ionescu.(2010b)}]{CI09m}
\bibinfo{author}{Cazacu, O.}, \bibinfo{author}{Ionescu., I.R.},
  \bibinfo{year}{2010}b.
\newblock \bibinfo{title}{Dynamic crystal plasticity: an eulerian approach}.
\newblock \bibinfo{journal}{Journal of Mechanics and Physics of Solids}
  \bibinfo{volume}{58}, \bibinfo{pages}{844--859}.
\bibitem[{Forest et~al.(2019)Forest, Mayeur and McDowell}]{Forest2019}
\bibinfo{author}{Forest, S.}, \bibinfo{author}{Mayeur, J.R.},
  \bibinfo{author}{McDowell, D.L.}, \bibinfo{year}{2019}.
\newblock \bibinfo{title}{Micromorphic Crystal Plasticity}.
  \bibinfo{publisher}{Springer International Publishing},
  \bibinfo{address}{Cham}.
\newblock pp. \bibinfo{pages}{643--686}.
\bibitem[{Franciosi(1985)}]{fra85}
\bibinfo{author}{Franciosi, P.}, \bibinfo{year}{1985}.
\newblock \bibinfo{title}{The concepts of latent hardening and strain hardening
  in metallic single crystals}.
\newblock \bibinfo{journal}{Acta Metall} \bibinfo{volume}{33},
  \bibinfo{pages}{1601--1612}.
\bibitem[{Glowinski and Le~Tallec(1989)}]{GlLT}
\bibinfo{author}{Glowinski, R.}, \bibinfo{author}{Le~Tallec, P.},
  \bibinfo{year}{1989}.
\newblock \bibinfo{title}{Augmented Lagrangian and Operator Splitting method in
  Non-Linear Mechanics}.
\newblock \bibinfo{publisher}{SIAM Studies in Applied Mathematics}.
\bibitem[{Hill(1998)}]{Hill1998-ae}
\bibinfo{author}{Hill, R.}, \bibinfo{year}{1998}.
\newblock \bibinfo{title}{The Mathematical Theory of Plasticity}.
\newblock \bibinfo{publisher}{Clarendon Press}.
\bibitem[{Hughes and Hansen(2018)}]{Hughes2018-ge}
\bibinfo{author}{Hughes, D.A.}, \bibinfo{author}{Hansen, N.},
  \bibinfo{year}{2018}.
\newblock \bibinfo{title}{The microstructural origin of work hardening stages}.
\newblock \bibinfo{journal}{Acta Mater.} \bibinfo{volume}{148},
  \bibinfo{pages}{374--383}.
\bibitem[{Hull and Bacon(2001)}]{hull2001introduction}
\bibinfo{author}{Hull, D.}, \bibinfo{author}{Bacon, D.J.},
  \bibinfo{year}{2001}.
\newblock \bibinfo{title}{Introduction to dislocations}.
\newblock \bibinfo{publisher}{Butterworth-Heinemann}.
\bibitem[{Hutchinson(1976)}]{hut76}
\bibinfo{author}{Hutchinson, J.W.}, \bibinfo{year}{1976}.
\newblock \bibinfo{title}{Bounds and selfconsistent estimates for creep of
  polycrystalline materials}.
\newblock \bibinfo{journal}{Proc. R. Soc. Lond. A} \bibinfo{volume}{348},
  \bibinfo{pages}{101--127}.
\bibitem[{Ionescu and Oudet(2010)}]{Ionescu2010-zc}
\bibinfo{author}{Ionescu, I.R.}, \bibinfo{author}{Oudet, {\'E}.},
  \bibinfo{year}{2010}.
\newblock \bibinfo{title}{Discontinuous velocity domain splitting in limit
  analysis}.
\newblock \bibinfo{journal}{Int. J. Solids Struct.} \bibinfo{volume}{47},
  \bibinfo{pages}{1459--1468}.
\bibitem[{Isp{\'a}novity et~al.(2014)Isp{\'a}novity, Laurson, Zaiser, Groma,
  Zapperi and Alava}]{Ispanovity2014-ra}
\bibinfo{author}{Isp{\'a}novity, P.D.}, \bibinfo{author}{Laurson, L.},
  \bibinfo{author}{Zaiser, M.}, \bibinfo{author}{Groma, I.},
  \bibinfo{author}{Zapperi, S.}, \bibinfo{author}{Alava, M.J.},
  \bibinfo{year}{2014}.
\newblock \bibinfo{title}{Avalanches in {2D} dislocation systems: plastic
  yielding is not depinning}.
\newblock \bibinfo{journal}{Phys. Rev. Lett.} \bibinfo{volume}{112},
  \bibinfo{pages}{235501}.
\bibitem[{Kachanov(1971)}]{Kachanov1971-he}
\bibinfo{author}{Kachanov, L.M.}, \bibinfo{year}{1971}.
\newblock \bibinfo{title}{Foundations of the theory of plasticity amsterdam}.
\bibitem[{Kok and Tortorelli(2002)}]{kok02}
\bibinfo{author}{Kok, S., B.A.J.}, \bibinfo{author}{Tortorelli, D.A.},
  \bibinfo{year}{2002}.
\newblock \bibinfo{title}{A polycrystal plasticity model based on the
  mechanical threshold.}
\newblock \bibinfo{journal}{Int.Journal of Plast.} \bibinfo{volume}{18},
  \bibinfo{pages}{715--741}.
\bibitem[{Lebensohn and Tom\'e(1993)}]{lebtome93}
\bibinfo{author}{Lebensohn, R.}, \bibinfo{author}{Tom\'e, C.N.},
  \bibinfo{year}{1993}.
\newblock \bibinfo{title}{A self-consistent anisotropic approach for the
  simulation of plastic deformation and texture development of polycrystals:
  Application to zirconium alloys}.
\newblock \bibinfo{journal}{Acta Metall. Mater.} \bibinfo{volume}{41},
  \bibinfo{pages}{2611--2624}.
\bibitem[{Maa{\ss} and Derlet(2017)}]{Maas2017-yx}
\bibinfo{author}{Maa{\ss}, R.}, \bibinfo{author}{Derlet, P.M.},
  \bibinfo{year}{2017}.
\newblock \bibinfo{title}{Micro-plasticity and recent insights from
  intermittent and small-scale plasticity}.
\newblock \bibinfo{journal}{Acta Mater.} .
\bibitem[{Marano et~al.(2019)Marano, G{\'e}l{\'e}bart and
  Forest}]{Marano2019-he}
\bibinfo{author}{Marano, A.}, \bibinfo{author}{G{\'e}l{\'e}bart, L.},
  \bibinfo{author}{Forest, S.}, \bibinfo{year}{2019}.
\newblock \bibinfo{title}{Intragranular localization induced by softening
  crystal plasticity: Analysis of slip and kink bands localization modes from
  high resolution {FFT-simulations} results}.
\newblock \bibinfo{journal}{Acta Mater.} \bibinfo{volume}{175},
  \bibinfo{pages}{262--275}.
\bibitem[{McDowell(2019)}]{McDowell2019-jl}
\bibinfo{author}{McDowell, D.L.}, \bibinfo{year}{2019}.
\newblock \bibinfo{title}{Multiscale modeling of interfaces, dislocations, and
  dislocation field plasticity}, in: \bibinfo{editor}{Mesarovic, S.},
  \bibinfo{editor}{Forest, S.}, \bibinfo{editor}{Zbib, H.} (Eds.),
  \bibinfo{booktitle}{Mesoscale Models: From {Micro-Physics} to
  {Macro-Interpretation}}. \bibinfo{publisher}{Springer International
  Publishing}, \bibinfo{address}{Cham}, pp. \bibinfo{pages}{195--297}.
\bibitem[{Pan et~al.(2019)Pan, Wu, Wang, Sun, Xiao, Ding, Sun and
  Salje}]{Pan2019-sl}
\bibinfo{author}{Pan, Y.}, \bibinfo{author}{Wu, H.}, \bibinfo{author}{Wang,
  X.}, \bibinfo{author}{Sun, Q.}, \bibinfo{author}{Xiao, L.},
  \bibinfo{author}{Ding, X.}, \bibinfo{author}{Sun, J.},
  \bibinfo{author}{Salje, E.K.H.}, \bibinfo{year}{2019}.
\newblock \bibinfo{title}{Rotatable precipitates change the scale-free to scale
  dependent statistics in compressed ti nano-pillars}.
\newblock \bibinfo{journal}{Sci. Rep.} \bibinfo{volume}{9},
  \bibinfo{pages}{3778}.
\bibitem[{Papanikolaou et~al.(2012)Papanikolaou, Dimiduk, Choi, Sethna, Uchic,
  Woodward and Zapperi}]{Papanikolaou2012-up}
\bibinfo{author}{Papanikolaou, S.}, \bibinfo{author}{Dimiduk, D.M.},
  \bibinfo{author}{Choi, W.}, \bibinfo{author}{Sethna, J.P.},
  \bibinfo{author}{Uchic, M.D.}, \bibinfo{author}{Woodward, C.F.},
  \bibinfo{author}{Zapperi, S.}, \bibinfo{year}{2012}.
\newblock \bibinfo{title}{Quasi-periodic events in crystal plasticity and the
  self-organized avalanche oscillator}.
\newblock \bibinfo{journal}{Nature} \bibinfo{volume}{490},
  \bibinfo{pages}{517--521}.
\bibitem[{Peirce and Needleman(1982)}]{peir82}
\bibinfo{author}{Peirce, D.~Asaro, R.}, \bibinfo{author}{Needleman, A.},
  \bibinfo{year}{1982}.
\newblock \bibinfo{title}{An analysis of nonuniform and localized deformation
  in ductile single crystals}.
\newblock \bibinfo{journal}{Acta Metall.} \bibinfo{volume}{30},
  \bibinfo{pages}{1087--1119}.
\bibitem[{Rice(1973)}]{rice87}
\bibinfo{author}{Rice, J..}, \bibinfo{year}{1973}.
\newblock \bibinfo{title}{Plane strain slip line theory for anisotropic
  rigid/plastic materials}.
\newblock \bibinfo{journal}{Journal of Mechanics and Physics of Solids}
  \bibinfo{volume}{21}, \bibinfo{pages}{63--74}.
\bibitem[{Roters et~al.(2010)Roters, Eisenlohr, Hantcherli, Tjahjanto, Bieler
  and Raabe}]{Roters2010-cw}
\bibinfo{author}{Roters, F.}, \bibinfo{author}{Eisenlohr, P.},
  \bibinfo{author}{Hantcherli, L.}, \bibinfo{author}{Tjahjanto, D.D.},
  \bibinfo{author}{Bieler, T.R.}, \bibinfo{author}{Raabe, D.},
  \bibinfo{year}{2010}.
\newblock \bibinfo{title}{Overview of constitutive laws, kinematics,
  homogenization and multiscale methods in crystal plasticity finite-element
  modeling: Theory, experiments, applications}.
\newblock \bibinfo{journal}{Acta Mater.} \bibinfo{volume}{58},
  \bibinfo{pages}{1152--1211}.
\bibitem[{Salman and Truskinovsky(2012)}]{SALMAN2012219}
\bibinfo{author}{Salman, O.}, \bibinfo{author}{Truskinovsky, L.},
  \bibinfo{year}{2012}.
\newblock \bibinfo{title}{On the critical nature of plastic flow: One and two
  dimensional models}.
\newblock \bibinfo{journal}{International Journal of Engineering Science}
  \bibinfo{volume}{59}, \bibinfo{pages}{219 -- 254}.
\bibitem[{Salman(2009)}]{Salmant}
\bibinfo{author}{Salman, O.U.}, \bibinfo{year}{2009}.
\newblock \bibinfo{title}{Modeling of spatio-temporal dynamics and patterning
  mechanisms of martensites by phase-field and Lagrangian methods}.
\newblock Ph.D. thesis. Paris 6.
\bibitem[{Salman and Baggio(2019)}]{Salman2019-no}
\bibinfo{author}{Salman, O.U.}, \bibinfo{author}{Baggio, R.},
  \bibinfo{year}{2019}.
\newblock \bibinfo{title}{Homogeneous Dislocation Nucleation in Landau Theory
  of Crystal Plasticity}. \bibinfo{publisher}{Wiley Online Library}.
\newblock pp. \bibinfo{pages}{1--24}.
\bibitem[{Salman and Truskinovsky(2011)}]{Salman2011-ij}
\bibinfo{author}{Salman, O.U.}, \bibinfo{author}{Truskinovsky, L.},
  \bibinfo{year}{2011}.
\newblock \bibinfo{title}{Minimal integer automaton behind crystal plasticity}.
\newblock \bibinfo{journal}{Phys. Rev. Lett.} \bibinfo{volume}{106},
  \bibinfo{pages}{175503}.
\bibitem[{Sparks and Maa{\ss}(2018)}]{Sparks2018-zp}
\bibinfo{author}{Sparks, G.}, \bibinfo{author}{Maa{\ss}, R.},
  \bibinfo{year}{2018}.
\newblock \bibinfo{title}{Nontrivial scaling exponents of dislocation
  avalanches in microplasticity}.
\newblock \bibinfo{journal}{Phys. Rev. Materials} \bibinfo{volume}{2},
  \bibinfo{pages}{120601}.
\bibitem[{Teodosiu and Sidoroff(1976)}]{TeodSid76}
\bibinfo{author}{Teodosiu, C.}, \bibinfo{author}{Sidoroff, F..},
  \bibinfo{year}{1976}.
\newblock \bibinfo{title}{A theory of finite elasto-viscoplasticity of single
  crystals}.
\newblock \bibinfo{journal}{Int J Engng. Sci.} \bibinfo{volume}{14},
  \bibinfo{pages}{165--176}.
\bibitem[{Teodosiu and Tabourot(1976)}]{TeodRafTab}
\bibinfo{author}{Teodosiu, C., R.J.}, \bibinfo{author}{Tabourot, L.},
  \bibinfo{year}{1976}.
\newblock \bibinfo{title}{A theory of finite elasto-viscoplasticity of single
  crystals}.
\newblock \bibinfo{journal}{Int J Engng. Sci.} \bibinfo{volume}{14},
  \bibinfo{pages}{165--176}.
\bibitem[{Weiss(2019)}]{Weiss2019-sg}
\bibinfo{author}{Weiss, J.}, \bibinfo{year}{2019}.
\newblock \bibinfo{title}{Ice: the paradigm of wild plasticity}.
\newblock \bibinfo{journal}{Philos. Trans. A Math. Phys. Eng. Sci.}
  \bibinfo{volume}{377}, \bibinfo{pages}{20180260}.
\bibitem[{Wilson(1954)}]{Wilson:a28461}
\bibinfo{author}{Wilson, A.J.C.}, \bibinfo{year}{1954}.
\newblock \bibinfo{title}{{Dislocations and Platic Flow in Crystals by A. H.
  Cottrell}}.
\newblock \bibinfo{journal}{Acta Crystallographica} \bibinfo{volume}{7},
  \bibinfo{pages}{384}.
\bibitem[{Wu et~al.(2019)Wu, Liu, Sun, Wang, Sun, Han, Kai, Luan, Liu, Cao, Lu,
  Cheng and Lu}]{Wu2019-lr}
\bibinfo{author}{Wu, G.}, \bibinfo{author}{Liu, C.}, \bibinfo{author}{Sun, L.},
  \bibinfo{author}{Wang, Q.}, \bibinfo{author}{Sun, B.}, \bibinfo{author}{Han,
  B.}, \bibinfo{author}{Kai, J.J.}, \bibinfo{author}{Luan, J.},
  \bibinfo{author}{Liu, C.T.}, \bibinfo{author}{Cao, K.}, \bibinfo{author}{Lu,
  Y.}, \bibinfo{author}{Cheng, L.}, \bibinfo{author}{Lu, J.},
  \bibinfo{year}{2019}.
\newblock \bibinfo{title}{Hierarchical nanostructured aluminum alloy with
  ultrahigh strength and large plasticity}.
\newblock \bibinfo{journal}{Nat. Commun.} \bibinfo{volume}{10},
  \bibinfo{pages}{5099}.
\bibitem[{Zhang et~al.(2020a)Zhang, Bian, Zhang, Liu, Weiss and
  Sun}]{Zhang2020-ve}
\bibinfo{author}{Zhang, P.}, \bibinfo{author}{Bian, J.J.},
  \bibinfo{author}{Zhang, J.Y.}, \bibinfo{author}{Liu, G.},
  \bibinfo{author}{Weiss, J.}, \bibinfo{author}{Sun, J.},
  \bibinfo{year}{2020}a.
\newblock \bibinfo{title}{Plate-like precipitate effects on plasticity of
  {Al-Cu} alloys at micrometer to sub-micrometer scales}.
\newblock \bibinfo{journal}{Mater. Des.} \bibinfo{volume}{188},
  \bibinfo{pages}{108444}.
\bibitem[{Zhang et~al.(2020b)Zhang, Salman, Weiss and
  Truskinovsky}]{zhang2020variety}
\bibinfo{author}{Zhang, P.}, \bibinfo{author}{Salman, O.U.},
  \bibinfo{author}{Weiss, J.}, \bibinfo{author}{Truskinovsky, L.},
  \bibinfo{year}{2020}b.
\newblock \bibinfo{title}{Variety of scaling behaviors in nanocrystalline
  plasticity}.
\newblock \href{http://arxiv.org/abs/2004.08579}{\tt arXiv:2004.08579}.
\bibitem[{Zhang et~al.(2017)Zhang, Salman, Zhang, Liu, Weiss, Truskinovsky and
  Sun}]{Zhang2017-cl}
\bibinfo{author}{Zhang, P.}, \bibinfo{author}{Salman, O.U.},
  \bibinfo{author}{Zhang, J.Y.}, \bibinfo{author}{Liu, G.},
  \bibinfo{author}{Weiss, J.}, \bibinfo{author}{Truskinovsky, L.},
  \bibinfo{author}{Sun, J.}, \bibinfo{year}{2017}.
\newblock \bibinfo{title}{Taming intermittent plasticity at small scales}.
\newblock \bibinfo{journal}{Acta Mater.} \bibinfo{volume}{128},
  \bibinfo{pages}{351--364}.

\end{thebibliography}


\end{document}